# System overview of the VLTI Spectro-Imager

L. Jocou[a], J.P. Berger[a], F. Malbet[a], P. Kern[a], U. Beckmann[b], D. Lorenzetti[c], L. Corcione[d], G. Li Causi[c], D. Buscher[e], J. Young[e], M. Gai[d], G. Weigelt[b], G. Zins[a], G. Duvert[a], K. Perraut[a], P. Labeye[f], O. Absil[a], P. Garcia[g], D. Loreggia[c], J. Lima[h], J. Rebordao[i], S. Ligori[d], A. Amorim[h], P. Rabou[a], J.B. Le Bouquin[j], C. Haniff[e], E. Le Coarer[a], P. Feautrier[a], G. Duchene[a], M. Benisty[a], A. Chelli[a], E. Herwats[ak], A. Delboulbé[a]

[a] Université J. Fourier, CNRS, Laboratoire d'Astrophysique de Grenoble, UMR 5571, BP 53, F-38041 Grenoble cedex 9, France;
[b] MPIfR : Max-Planck Institute for Radioastronomy, Bonn, Germany;
[c] INAF/OAR: INAF/Osservatorio di Astrofisica di Roma, Italy;
[d] INAF/OATo: INAF/Osservatorio Astrofisico di Torino, Italy;
[e] Cavendish: Cavendish Laboratory of University of Cambridge, UK;
[f] CEA/LETI: Laboratoire d'electronique et de technologie de l'information, Grenoble, France;
[g] CAUP: Center for Astrophysics of University of Porto, Portugal;
[h] FCUL: Faculdade de Ciências da Universidade de Lisboa, Portugal;
[i] INETI: Instituto Nacional de Engenharia, Tecnologia e Inovacco, Lisboa, Portugal;
[j] ESO/S: European Southern Observatory, Santiago, Chile;
[k] IAGL: Institut d'Astrophysique et de Géophysique de Liège, Belgium;

**ABSTRACT**

The VLTI Spectro Imager project aims to perform imaging with a temporal resolution of 1 night and with a maximum angular resolution of 1 milliarcsecond, making best use of the Very Large Telescope Interferometer capabilities. To fulfill the scientific goals (see Garcia et. al.), the system requirements are: a) combining 4 to 6 beams; b) working in spectral bands J, H and K; c) spectral resolution from R= 100 to 12000; and d) internal fringe tracking on-axis, or off-axis when associated to the PRIMA dual-beam facility.

The concept of VSI consists on 6 sub-systems: a common path distributing the light between the fringe tracker and the scientific instrument, the fringe tracker ensuring the co-phasing of the array, the scientific instrument delivering the interferometric observables and a calibration tool providing sources for internal alignment and interferometric calibrations. The two remaining sub-systems are the control system and the observation support software dedicated to the reduction of the interferometric data.

This paper presents the global concept of VSI science path including the common path, the scientific instrument and the calibration tool. The scientific combination using a set of integrated optics multi-way beam combiners to provide high-stability visibility and closure phase measurements are also described. Finally we will address the performance budget of the global VSI instrument. The fringe tracker and scientific spectrograph will be shortly described.

## 1. INTRODUCTION

The VLTI Spectro Imager (VSI) will be capable to combine up to 6 telescopes allowing imaging. It aims to provide ESO community with spectrally-resolved near-infrared images at angular resolutions down to 1.1 milli-arcsecond and spectral resolutions up to R = 12000 in the J, H and K spectral bands (see Malbet et. al.[1] and Garcia et. al.[2] ). The instrument includes its own fringe tracker and wavefront control in order to reduce the constraints on the VLTI infrastructure and maximize the scientific return. Fainter targets up to K = 10 can be imaged without requiring a brighter nearby reference object; fainter targets can be accessed if a suitable off-axis reference is available thanks to the dual feed operation mode. The concept presented in this paper is the result of a phase A study aiming to demonstrate that no major issues has to be considered in the next phase. VSI will be developed to combine 4 telescopes in its first phase before reaching its most evolved version combining 6 telescopes. Although VSI is developed to combine 6 telescopes, its design allow an extension to an 8 telescopes operation.

## 2. VSI OVERVIEW

The VSI system is composed of the following sub-systems (Figure 1):

**1. Common Path (CP):** ensures that the VLTI incoming light is properly distributed between the fringe tracker and the scientific instrument.

**2. Fringe Tracker (FT):** ensures the best possible level of array coherencing/cophasing by monitoring in real time optical path fluctuations;

**3. Scientific Instrument (SI):** delivers the interferometric observables that will lead to the final image reconstruction. Its function is to record the interferometric signal gathered on scientific targets and calibration sources (both on sky and in lab).

**4. Calibration-Alignment Tools (CAT):** provides coherent and uncoherent sources to make alignment and internal interferometric calibrations of both the FT and SI during integration or operation phases;

**5. Control system:** computers, software and electronics for controlled functions such as motors, piezo devices, detectors, sensors and others active systems in VSI;

**6. Observation support software:** software package dedicated to the reduction of the interferometric data aiming to provide up to reconstructed images.

The control system and the observation software will not be described in this paper.

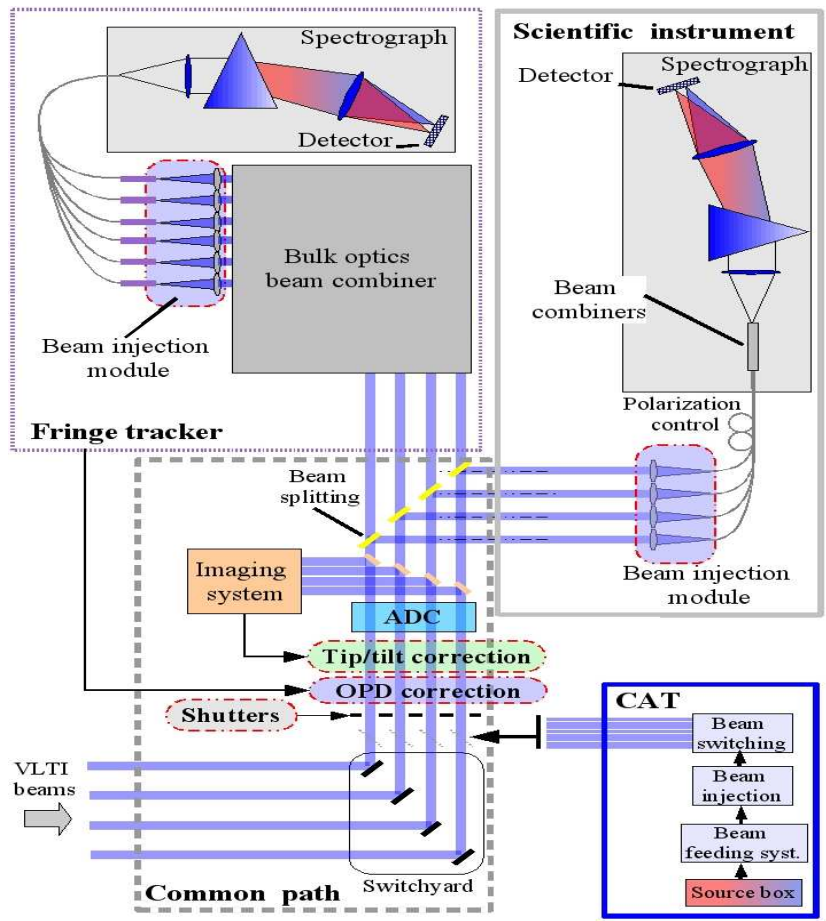

Figure 1: Description of the VSI system.

## 3. VSI SYSTEM DESCRIPTION

A global implementation is proposed in Figure 2. This considers the use of three independent optical tables leading to an overall dimension of the 4 m x 1,6m.

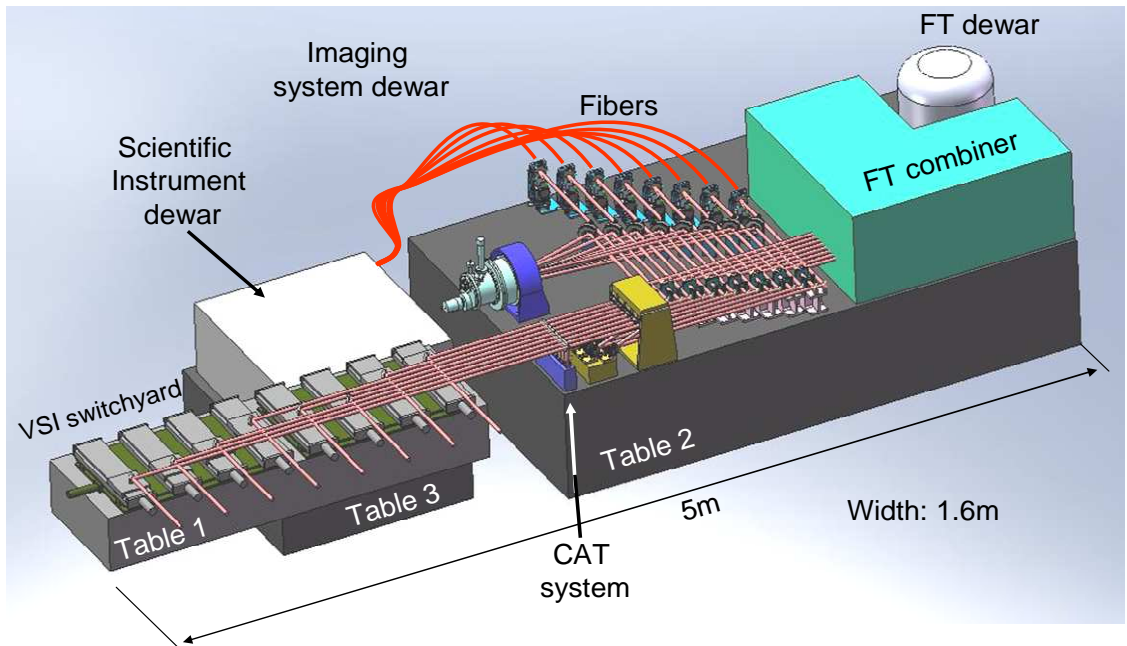

Figure 2: Global implementation of the VSI instrument. Three tables are used to implement the system. The common path is spread over the tables 1 and 2 up to the fiber injection (in red on the figure). The fringe tracker is located on table 2 while the scientific instrument dewar is located on the table 3. The CAT feeding optics is located on the table 2.

## 4. COMMON PATH CONCEPTUAL DESIGN

This assembly aims at providing suitable beams to FT and SI. It integrates correcting devices for optical path and image stabilization. It includes the following functions:

- Beam switchyard;
- Correct the dynamic tip-tilt variation of the beams;
- Image the target to improve alignment operation;
- Distribute the beams between fringe tracker and scientific instrument including the dual feed operation.

VSI will be a facility able, from the beginning of its operation, to perform fringe tracking on an off axis bright stars using the dual feed PRIMA mode. In this mode VSI and especially the common path, will have to manage 8 beams i.e. 4 beams dedicated to the reference and 4 beams dedicated to the star, (see

 Figure **3**).

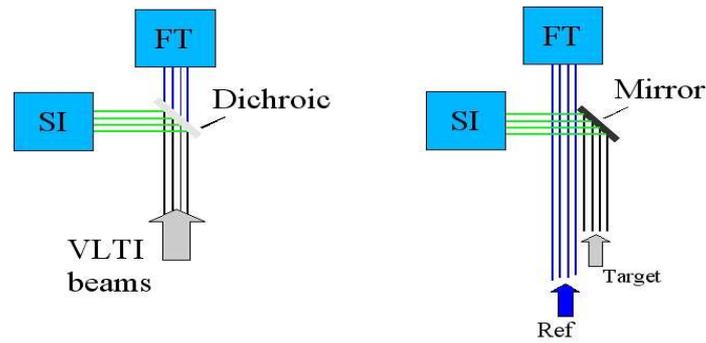

Figure 3: Description of the dual feed PRIMA mode: On the left, the 4 or 6 VLTI beams are distributed between the scientific instrument and the fringe tracker by a dichroic. In dual feed mode, 4 beams coming from the brightest star, are feeding directly the fringe tracker while the others 4 beams are distributed in the science path, by a folding mirror.

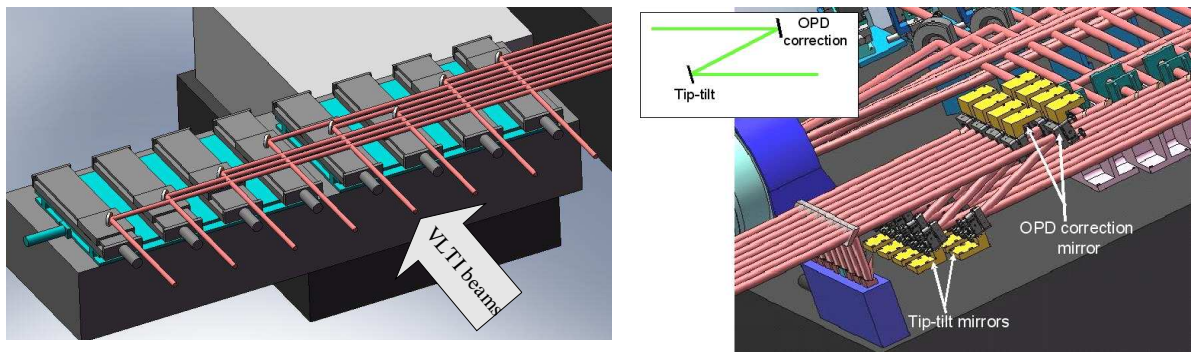

Figure 4: Description of the common path. Left: zoom of the table 1 supporting the beam switchyard. Right: Detail of the optical layout. The beams are folded once by the tip-tilt mirror, and a second time by the OPD mirror controlled by the fringe tracker.

### 4.1 Switchyard

As we can see Figure 4, the first optical table supports an internal switchyard. The fringe tracking operating in boostrapping mode, the switchyard is required to redirect any incoming beam toward any arm of the fringe tracker beam combiner. In our concept each mirror is held on a motorized stage to address the beams in the convenient arms of VSI.

### 4.2 Tip-tilt control and imaging system

Since VSI is using single-mode fibers, the tip-tilt variations could highly affect the coupling of the light into the instrument. Wavefront measurement campaign currently in progress at Paranal in the VLTI lab will drive the definition of the specifications for this subsystem. Our design however includes a tip-tilt mirror aiming at optimizing and stabilizing the coupling of the VLTI beams in the fibers.

We will implement an *imaging system* used both as a tip-tilt sensor and as a mean to adjust the pointing position of the stellar object on the fiber. A set of corner cubes is therefore included in the common path to image simultaneously the target and fiber ends.

The common path includes also a set of mirrors controlled by the fringe tracker to correct the opd fluctuations. The tip-tilt mirrors and the opd correction mirrors fold in a 'Z' shape as described Figure 4.

## 4.3 Beam distribution

The last element of the common path is a beam splitter device dedicated to split the light between the fringe tracker and scientific instrument according to the Table 1. A fourth configuration implements a set of mirrors to fold 4 beams and feed the scientific instrument for the dual feed operation (see Figure 3).

| Instrumental configuration | Scientific instrument operating band | Fringe tracker operating band |
|---|---|---|
| 1 | J or H | K |
| 2 | K | H |
| 3 | K | K |

Table 1: Beam distribution between the FT and the scientific instrument

The beam splitter concept is described Figure 5. It uses a single dichroic beam splitter to manage the instrumental configurations 1 and 2 of the Table 1. The switching between these two modes is obtained by inserting the mirror M3 in the optical layout. In this configuration the dichroic is specified to have the K band transmitted and I-J-H bands reflected.

This concept reduces the number of dichroic beamsplitters. To manage the instrumental configuration 3, the M3 mirror could be replaced by a simple beam splitter optimized in K band.

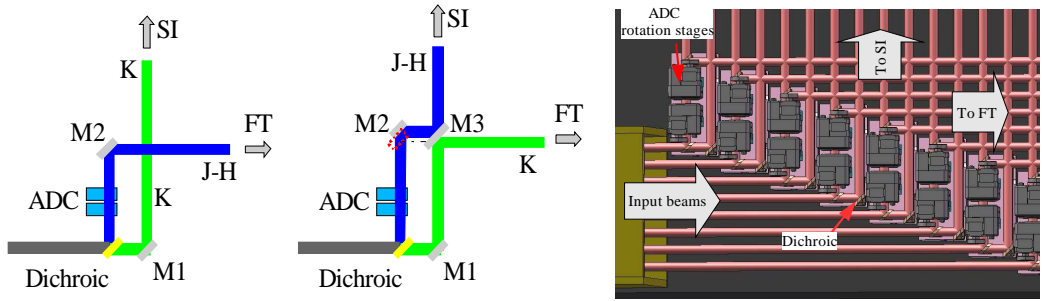

Figure 5: Description of the beam splitting device. Left: strategy proposed to distribute the light between fringe tracker and scientific instrument. The switching between the two configurations is obtained by inserting the M3 mirror. This concept allows, whatever the configuration, to implement the ADC device only in the J-H beam. Right: implementation of the beam splitting device. The grey part is the ADC rotation stage.

## 5. SCIENTIFIC INSTRUMENT CONCEPTUAL DESIGN

The functions complied by the scientific instrument are:
- Collect the VLTI beams to feed the beam combiner fibers;
- Combine the beam to deliver observables;
- Provide spectral dispersions;
- Allows the recording of the interferometric signals.

## 5.1 VLTI beam collection

This function is made by the injection modules injecting the 18mm VLTI beams into the fibers feeding the integrated optics beam combiner. The concept of this module is based on an off-axis parabolic mirror feeding light into the suitable fiber according to the operating wavelength (see Figure 6). Five degrees of freedom on the fiber head for the (x, y, z) motorized translations, and on the whole assembly for the tip-tilt are required to optimize light coupling in each fiber. A global translation allows both to adjust OPD and to correct the differential longitudinal atmospheric dispersion appearing between the fringe tracker and the scientific instrument for large zenithal angle.

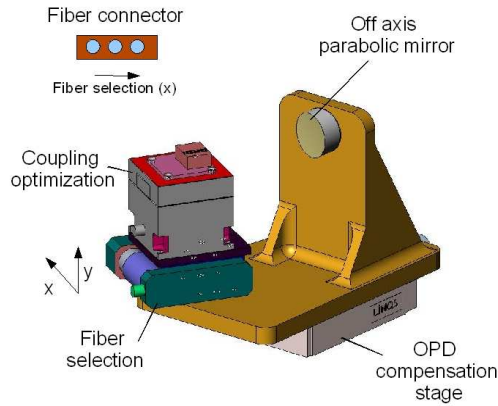

Figure 6: Description of the injection module. One module integrates 3 three fibers i.e. one per spectral band (see upper part of the figure). By selecting the fiber to feed, the module select the operating spectral band of the scientific instrument.

## 5.2 Beam combination

In the VSI scientific instrument the beam combination relies on several integrated optics (hereafter I.O) chips which ensure the beam combining function. In planar optics, it is difficult to manufacture IO chips optimized over the whole spectral range. Therefore in VSI one IO chip is optimized for one spectral band in such a way three distinct IO chips will be implemented to cover J, H and K bands. Among all the beam combination concepts, our system study has pointed out that in the four telescopes configuration, a pairwise ABCD beam combiner was the best compromise between performance and number of used pixels (Benisty et al. [3] ). This combination concept has thus been designed in collaboration with our partner LETI[aA]. The resulting circuit is presented Figure 7. The combination chip has 4 inputs and 24 outputs (6 baselines and 4 A,B,C,D samples per baseline) (see Figure 7). The use of achromatic phase shifters allows different coherent phase states to be sampled. No temporal modulation is then anymore required for this concept. We have shown that due to the pairwise combination scheme the data reduction process can be turned into a matrix inversion issue. With proper calibration implementation, one can retrieve photometry, visibilities and phases from a single measurement of 24 intensities.

The laboratory measurements performed on the 4T-ABCD demonstrated a contrast larger than C=95%. The transmission has been measured at 65% in H band. A new design has been proposed and is under manufacturing to optimize the transmission to reach a value of 70% in J and H. The transmission in K should reach 60%.

Single-mode fibers ensure the guiding of the light from the beam injection module to the beam combiner. The coupling between the fibers and a IO beam combiner is ensured by cementing. The length of the fibers will be equalized to reduce the chromatic dispersion effects. A dedicated bench is available at LAOG to manufacture equalized fiber bundle with accuracy down to 20 µm.

---

[a] http://www-**leti**.cea.fr

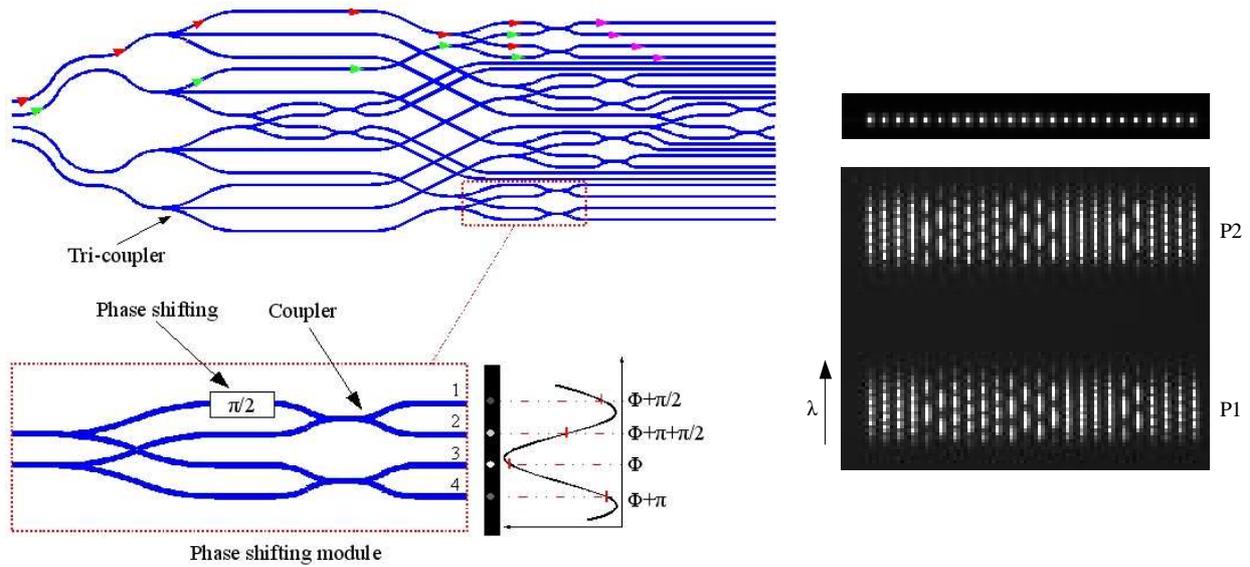

Figure 7: Design of the 4-way pairwise ABCD beam combiner. Left bottom: view of the circuit. Left bottom: principle of the phase shifting function allowing to sample the fringes with 4 phase states. Right: Image of the integrated optics beam combiner output. Top: image without dispersion showing the 24 outputs. Bottom: image of a laser diode with a low spectral disperser and a Wollaston prism splitting the two polarizations.

System analysis showed that the IO beam combiner should be cooled down to 233 K to limit the background emission. The set of IO beam combiners will be therefore implemented in the spectrograph dewar. The mechanical support of the IO beam combiner will be studied to contribute to the cooling control of the IO chip.

Preliminary tests have been performed to validate the performance of such cemented connectorizations between fibers and IO chip at a temperature of about 230 K. No degradation has been observed, but ageing cycles must be done in the coming month on the assembly.

Finally, the system includes a control of the polarization to ensure stability and contrast level. We will implement the combination of a perfectly equalized combiner (fiber + integrated optics (IO) combiner) and passive polarization compensation based on Lefebvre loops already used by ESO/VLTI (VINCI), FLUOR (IOTA).

For the 6 way beam combination, Lebouquin et. al.[4] has demonstrated that the most efficient design in terms of signal-to-noise ratio on the visibility measurements is a concept based on a multi-axial all-in-one combiner which allows fringes to be non-redundantly spatially encoded. The concept of the combiner is presented in Figure 8. The manufacturing of the 6-way is currently under way. Its characterization is foreseen by the end of 2008.

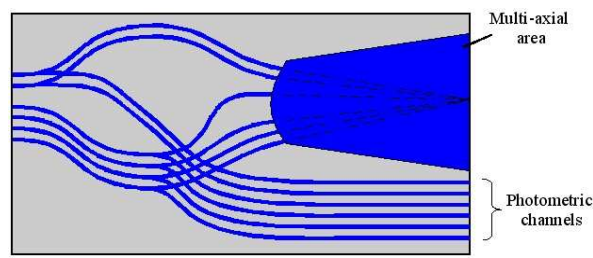

Figure 8: Description of the 6 telescopes multi-axial beam combiner

## 5.3 Spectrograph and detector

We make here a brief description of the spectrograph. A complete description of the spectrograph is given in Lorenzetti et. al.[5].

An overall optical layout of the spectrograph is given in Figure 9. The optical concept is based on three collimators each optimized for a specific band J, H and K and a common camera. The selection of the band is then ensured by a movable flat mirror keeping the beam combiners, their feeding fibers and the collimators in fixed positions. Four dispersive elements are supported on a rotating turret: a prism giving a resolution of 100 while three gratings are used to reach the resolutions 2000, 5000 and 12000.

The opto-mechanical part of the spectrograph is cooled down to reduce the thermal background emission. As said before the spectrograph will implement IO beam combiner substrates to be cooled at 233°K. The design foresees to integrate the scientific detector in the spectrograph dewar to optimize stability. The optics layout is mounted on an aluminum optical bench that is also the back-plate of the nitrogen vessel of the cryostat to produce a uniform cooling of the optical components. The whole optical bench is then enclosed in a radiation shield. The current footprint of the spectrograph is about 1x1x0.4m.

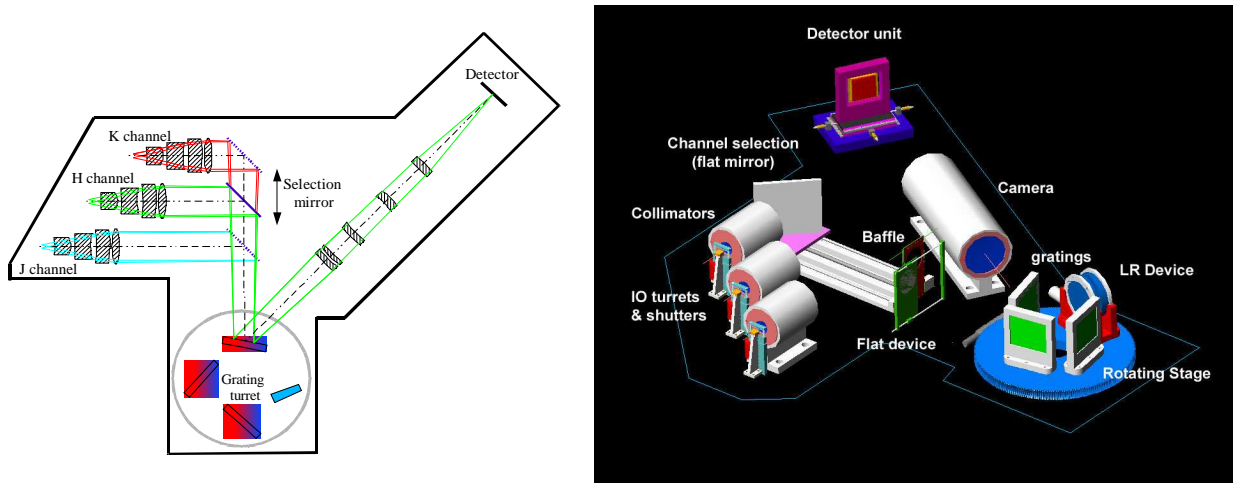

Figure 9: Scientific spectrograph implementation. Left: global view. Right: opto-mechanical implementation of the spectrograph (the table is not presented here).
.

The detector chosen for the VSI scientific instrument is a Hawaii-2RG coupled with the new NGC/ESO detector controller (NGC). The Hawaii-2RG, consisting of 32 strips of 64 pixels, read in parallel, is specifically adapted to have a fast read out of the 24 outputs of the beam combiner. The features of the detector are the following:

- Detector format required: 1k x 2k with a pixel size of 18 microns
- Readout noise: < 15 electrons
- Available readout modes: Correlated double sampling, Uncorrelated and Up-the-ramp

# 6.  FRINGE TRACKER CONCEPTUAL DESIGN

The FT will have to operate in three distinct configurations:

- "**Co-phasing**" configuration in which the OPD fluctuations are actively compensated at high speed so that the level of stability of the science combiner fringes is sufficient to allow long exposure. This configuration is consequently related to bright sources for the FT.

- ``**Coherencing**" configuration in which the OPD fluctuations are compensated in real time but at a lower speed and lower accuracy. It ensures that the loss in fringe contrast due to temporal coherence effects is not too large. The fringes are not anymore stabilized, and long exposures are not anymore achievable. This configuration is required for tracking on fainter sources.

- ``**Fringe acquisition**" configuration in which a finite region of OPD space is scanned to find the fringe coherence envelope in the presence of atmospheric and instrumental delay uncertainties.

A full description of the fringe tracker concept is given in Corcione et. al [6].

The fringe tracker concept is based on a bulk optics beam combiner with a temporal ABCD coding of the fringe to ensure the cophasing operation and a spectral dispersion of the fringes to allow coherencing mode. The Fringe tracker layout is presented Figure 10. It is developed to manage the fringe tracking on 6 beams. The beam combination is done before the spatial filtering: the VLTI beams are entering from the left before being splitted in two parts. Each beam is then combined with the nearest neighbor beam to produce the fringes. An additional optics is used to combine the beams 1 and 6 and therefore close the telescope ring and insure a redundant beam combination. At the output of the second dichroic stage, a pair of beams emerges in phase opposition (the so-called AC pair of measurements). A temporal modulation of the mirror bank (see on the Figure 10 is applied to obtain the two other states of phase BD and thus access to the quadrature outputs.

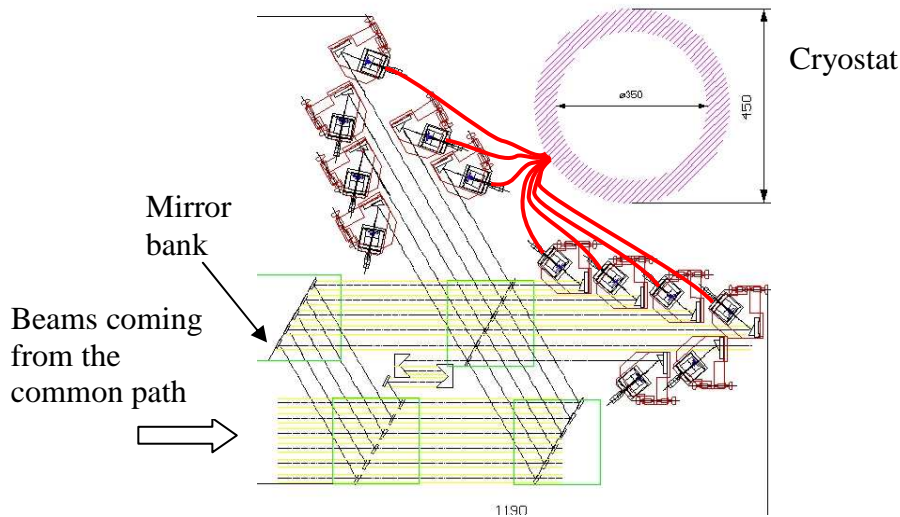

Figure 10: Optical layout of the fringe tracker. Only 7 fibers are represented on this figure.

Each combined beam is then collected to be injected in single-mode fibers mainly used to reduce the thermal background in K band. The current design of the FT foresees to use a unique single mode K fiber for both operational H and K bands.

At the other end, the fibers are integrated in the dewar including a cooled low resolution spectrograph allowing the fringes to be dispersed over 3 to 5 pixels of the detector. The detector chosen for the FT is a 1K x 1K Rockwell HAWAII-1 FPA featuring a readout noise between 5 to 15 electrons according to the readout mode.

The mirror allowing correcting the OPD fluctuations in real time is implemented in the common path assembly (see section 4.2)

## 7. CALIBRATION AND ALIGNMENT TOOL (CAT)

This tool feeds the VSI beams with coherent and incoherent sources for the calibration and alignment of both fringe sensor and scientific instrument. Its functions are:

- Provide sources for interferometric, spectral calibration and alignment
- Produce 6 coherent beams switchable to replace VLTI beams

This sub-system allows, for interferometric and calibration purposes, producing internal fringes to measure the contrast level for each pair of telescopes and thus estimate the scientific instrument performance during assembly, integration, tests and operation. A schematic view of the concept proposed for the CAT is shown Figure 11. The CAT produces 4 coherent channels (of almost identical flux) to feed the 4 arms of the FT and SI beam combiners.

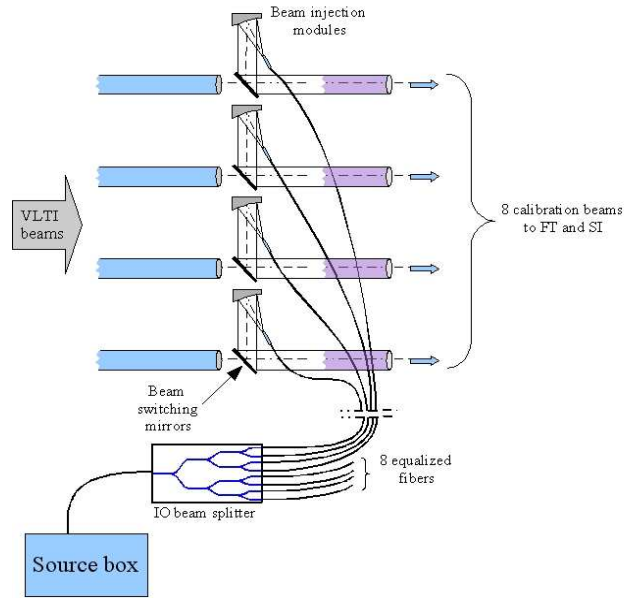

Figure 11: Description of the CAT concept

The CAT is fed by a box integrating the required sources such as HeNe laser, laser diode, or spectral calibration and neutral density filters. The spectral calibration is made with a set of Fabry-Perot devices that provide at least 10 absorption lines on the detector whatever the resolution. To fulfill with this requirement, two Fabry-Perot will be required. These etalons will be made of zerodur material to guarantee the level of stability required.

A switching device is implemented to substitute the VLTI beams by the CAT beams. To ensure a calibration capability as stable as possible, we implement an integrated optic device which allows the light of the source box to be splitted into 8 outputs equalized in flux. Each output of the IO beam splitter is coupled to one fiber dedicated to one arm of VSI. Like for the SI, three beam splitters will have to be used to cover the three spectral bands. Suitable equalized fibers must be connected to these beam splitters.

Finally to reproduce the VLTI beams, we use a module analog to the beam injection one of the SI but used up side down. The fibers coming from the IO beam splitter feed the 8 CAT arms.

## 8. EXPECTED PERFORMANCES

A preliminary error and performance budget has been done during the phase A study. The Table 2 presents the transmission budget of the scientific path of VSI. The Figure 12 describes the error budget defined for the contrast level taking into account the wavefront error, polarization and chromatic OPD effects.

These performances allow to reach a limiting magnitude in H up to K=10 with UT (K=7 with AT) if we consider a precision of 1% in squared visibility and 0,01 radian in closure phase (integration time: 100s, Fried parameter: 0.7").

| Sub-system | λ=1,25μm | λ=1,6μm | λ=2,1μm | λ=2,3μm |
|---|---|---|---|---|
| Common path optics | 0,72 | 0,72 | 0,72 | 0,72 |
| SI injection | 0,75 | 0,75 | 0,75 | 0,75 |
| SI beam combiner | 0,63 | 0,64 | 0,50 | 0,29 |
| Filters | 0,85 | 0,87 | 0,87 | 0,9 |
| Spectrograph optics | 0,73 | 0,73 | 0,72 | 0,65 |
| Gratings | 0,8 | 0,8 | 0,8 | 0,8 |
| Detector QE | 0,51 | 0,55 | 0,6 | 0,6 |
| **Total** | **0,083** | **0,093** | **0,077** | **0,042** |

Table 2: Transmission budget of the scientific instrument path

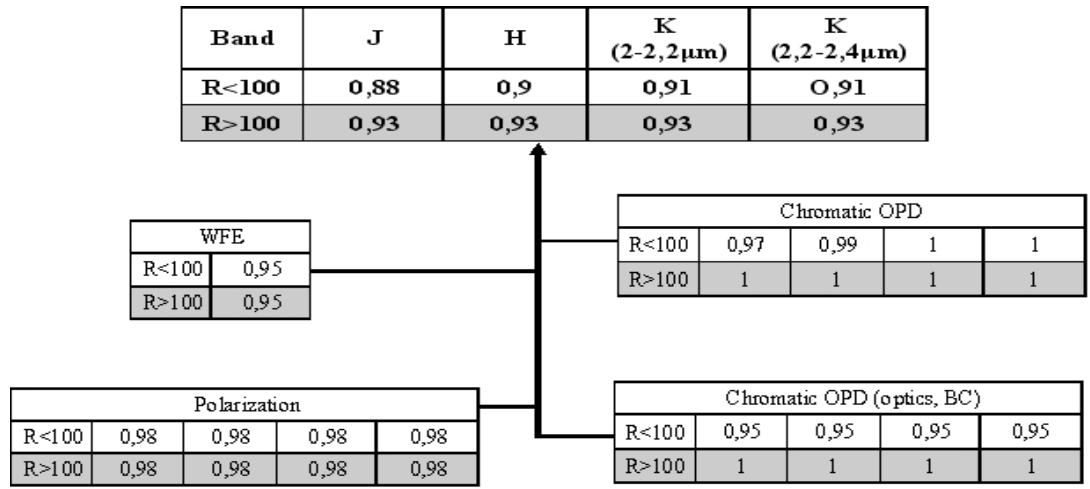

Figure 12: Contrast error budget. The budget is made in the three spectral bands (columns) and for two cases in spectral resolution: higher or smaller than R=100 (rows)

## 9. CONCLUSIONS AND PERSPECTIVES

The VSI phase A study did not reveal any show stopper on the development of the instrument. VSI should therefore begin its phase B in mid 2009 and should be integrated at VLTI in 2015. The coming years will be dedicated to validate and optimize the concept of VSI with the interferometric simulator available at LAOG.